\documentclass[%
 reprint, twocolumn, 
 amsmath,amssymb,
 aps, physrev,
floatfix
]{revtex4-2}

\usepackage{graphicx}
\usepackage{dcolumn}
\usepackage{bm}
\usepackage[normalem]{ulem}
\usepackage[colorlinks=true]{hyperref}
\usepackage[mathlines]{lineno}

\usepackage{hyperref}
\usepackage{xcolor}
\usepackage{physics}

\begin{document}
		
\title{Qubit reset beyond the Born-Markov approximation: optimal driving to overcome polaron formation}
\author{Carlos Ortega-Taberner}
\author{Eoin O'Neill}
\author{Paul Eastham}
\affiliation{School of Physics, Trinity College Dublin, Dublin 2, Ireland}
\affiliation{Trinity Quantum Alliance, Unit 16, Trinity Technology and Enterprise Centre, Pearse Street, Dublin 2, Ireland}

\date{\today}
\begin{abstract}
Qubits are typically reset into a known state by coupling them to a low-temperature environment. When treated in the Born-Markov approximation such couplings produce exponential relaxation to equilibrium, giving high reset fidelities limited only by temperature. We investigate qubit reset beyond this approximation, using numerically exact tensor network methods and the time-dependent variational principle, focussing on a spin-boson model describing a transmon qubit coupled to a resistor. Beyond the Born-Markov approximation the reset fidelity becomes limited by the buildup of system-environment correlations which corresponds to the formation of a polaron. We implement numerical optimal control to find time-dependent qubit Hamiltonians which overcome this limitation by steering the dynamics of the correlated system-environment state. The optimal controls becomes more effective when the environment is filtered to span a smaller spectral range, and remain effective when the multilevel nature of the transmon is considered. A related paper [C. Ortega-Taberner, E. O'Neill and P. R. Eastham, arXiv:XXXX.XXXX] addresses the complementary case of control via a time-dependent system-environment coupling. Our results show how limitations on reset speed and fidelity can be overcome, and how time-dependent driving can steer system-environment correlations and reverse polaron formation.
\end{abstract}

\maketitle

\section{Introduction}

The study of quantum systems coupled to an environment, i.e., open quantum systems, is increasingly important in many areas of physics and chemistry~\cite{nitzan2006,leggett1987,Uthailiang2026SpinBosonBio}. While the coupling can limit quantum devices and processes by producing decoherence and dissipation, it can also generate capabilities by allowing thermalization, dissipative state preparation, and the emergence of classical behavior. Open systems are most commonly studied for weak coupling to spectrally flat environments, which permits the Born-Markov approximation to be used to obtain time-local equations-of-motion for the system density matrix.  Although other possibilities have been considered, most qubits operate in regimes where these approximations are reasonable. However, very high fidelities have been achieved for gates, state preparation, and readout, and still further improvements are sought. Modelling systems to this accuracy requires treatments that capture physics beyond the Born-Markov approximation.

In this paper and a related Letter~\cite{shortpaper} we investigate the process of dissipative qubit reset, achieved by coupling a qubit to a low-temperature bath such that it relaxes to its ground state~\cite{reed2010,jones2013,tuorila2017,egger2018,magnard2018,basilewitsch2019,basilewitsch2021,zhou2021,sevriuk2022,aamir2023,diniz2023,yoshioka2023,yuan2023,gautier2024,shubrook_numerically_2025}. While the spin-boson model we use describes many systems~\cite{nitzan2006,leggett1987,Uthailiang2026SpinBosonBio}, we specifically consider the case of a superconducting transmon qubit coupled to a resistor.  Within the Born-Markov approximation the qubit will relax to equilibrium with the environment, implying residual thermal populations $\sim 10^{-5}$ for typical transmons. However, simulations beyond the Born-Markov approximation show a saturation of the population at a larger, coupling-dependent value, which implies a trade-off between speed and fidelity~\cite{tuorila2019}. We use tensor network methods~\cite{fux2024,link2024,strathearn2018} to confirm this observation, and show that it is due to the formation of a polaron
~\cite{silbey1984,xu2016} as the qubit relaxes towards the pure state. We then show how optimal control~\cite{butler2024,fux2021} can be used to remove this system-environment entanglement, allowing for fast, high-fidelity qubit reset. In our Letter we focus on control via the system-environment coupling, while the present paper explores control via a time-dependent qubit Hamiltonian. We investigate, furthermore, the role of the environment's spectral density, show how the optimal reset can be further improved using filters, and demonstrate that the optimal protocols remain effective when the multilevel nature of the transmon is considered. 

The formation of the polaron is a known feature of the spin-boson model, where the spin is entangled with coherent states from the bath, capturing the classical behavior of the environment \cite{bera2014}. The polaron was described by Landau and Pekar \cite{landau1933,landau1948}, who coined the term to describe a quasiparticle formed by an electron dressed by surrounding phonons. Polarons are known to influence the electrical and optical properties of many materials \cite{franchini2021}, and participate in a multitude of solid state phenomena such as high temperature superconductivity \cite{alexandrov1989,mott1993, franchini2021,zhang2023} or colossal magnetoresistance \cite{franchini2021}. The concept has also been extended to other types of systems. Particularly relevant to the spin-boson model is the case of localized excitons in quantum dots \cite{akimov2006,hohenester2007,manson2016,denning2020,preuss2022} and impurities in two-dimensional materials \cite{yu2024,vannucci2024,mitryakhin2024,steinhoff2025}, which are used as single-photon sources whose efficiency is known to be affected by polaron formation. Polaron formation also plays a major role in photochemistry~\cite{nitzan2006,Uthailiang2026SpinBosonBio}, where it accounts for the changing nuclear potential energy surface among electronic states. Our work thus has applications beyond qubit reset, by demonstrating how optimal control can be used to manipulate system-environment correlations and reverse polaron formation. 

The remainder of this paper is structured as follows. In Sec.~\ref{sec:resetandpolaron} we describe the model and discuss the limitations on reset arising in the standard protocol, with a fixed qubit frequency, and the connection to polaron formation. In Sec.~\ref{sec:optfreqcont} we use optimal control with the process tensor variant~\cite{fux2021} of the TEMPO algorithm~\cite{strathearn2018}  (PT-TEMPO) to find time-dependent Hamiltonians which mitigate these limitations. We use the polaron ansatz and the time-dependent variational principle to analyze the dynamics under these Hamiltonians, and show how they effectively destroy the polaron which forms during the initial stage of the reset. In Sec.~\ref{sec:tailoring} we investigate the effect of filtering the environment and show that this further improves the performance, while Sec.~\ref{sec:multilevel} shows that these protocols remain effective once the multilevel nature of the transmon is considered. Sec.~\ref{sec:concl} gives our conclusions and outlook. 

\section{Qubit reset and the polaron}
\label{sec:resetandpolaron}

We consider a superconducting qubit coupled to an environment, modelled by the spin-boson Hamiltonian
\begin{align}\label{eq:sb}
    H = \frac{\omega_q(t)}{2}\sigma_x + \sum_k \omega_k b_k^\dagger b_k + \frac{\sigma_z}{2}\sum_k g_k (b_k^\dagger + b_k),
\end{align}
where $\omega_q(t)$ is the qubit frequency, with typical values $\omega_q/2\pi \in  [4,7]$ GHz for transmons, and $g_k$ is the coupling to each bath mode. The effect of the environment is characterized by the spectral density, for which we take an Ohmic form with an exponential cutoff,
\begin{align}\label{eq:j1}
    J(\omega) =& \sum_k \abs{g_k}^2 \delta(\omega-\omega_k) \nonumber \\
    =& 2\alpha \omega e^{-\omega/\omega_c}.
\end{align} To produce the dissipation needed for reset the cutoff frequency, which corresponds to the peak of the spectral density, should be similar to the qubit frequency; we take $\omega_C/2\pi =  5$ GHz. $\alpha$ is a dimensionless coupling strength, which controls the rate of dissipation in the weak-coupling limit. 

In the following we consider a dissipative reset process where the qubit-environment coupling is time-independent during the reset. However, it should be noted that the coupling must be switched on at the start of the reset, and off afterwards. These switches, which we assume to be instantaneous, are required for fast reset, since otherwise the environment would interfere with normal operation of the qubit. The effect of a time-dependent coupling during the reset is considered elsewhere~\cite{shortpaper}.

The dynamics of the spin-boson Hamiltonian, Eq.~\eqref{eq:sb}, can be computed exactly using the PT-TEMPO algorithm~\cite{fux2021} as implemented in OQuPy~\cite{fux2024}. This algorithm operates in discrete time, and constructs the influence functional describing the effect of the environment as a tensor network, which is contracted to form a matrix-product operator with low bond dimension. This matrix product operator representation of the process tensor can be used to efficiently compute the dynamics for different time-dependent qubit Hamiltonians. To further increase the computational efficiency we use the infinite network contraction scheme~\cite{link2024} as well as analytical forms for the environment correlation functions (see Appendix~\ref{appendix:influence}).

In Fig.~\ref{fig:fig1} we show the residual excited state population for a standard reset with a time-independent qubit splitting. We take this splitting $\omega_q(t)=\omega_q^0=2\pi\times 5 $ GHz to correspond to the peak in the spectral density to maximize the dissipation rate. The initial state is a product state of the qubit and bath, with the qubit in the maximally mixed state, and the bath in a thermal state, which we take to be at zero temperature. The results are shown as a function of the protocol duration, $t_f$. Note that for this time-independent protocol this is the same as the population as a function of time, but that is not the case generally. 

The three dashed curves show populations for three different coupling strengths. For the two smallest values of the coupling strength the initial relaxation is exponential, as predicted by the Born-Markov approximation. The same behavior occurs for the largest coupling except for a short period at the beginning. Following the exponential relaxation, however, all three curves show the excited state population saturating to a non-zero value~\cite{tuorila2019}, contrary to the prediction of Born-Markov theory that the qubit thermalizes with the zero temperature bath.

We find~\cite{shortpaper}  that this saturation behavior corresponds to the formation, at the end of the dynamics, of the polaron state \cite{silbey1984,xu2016}
\begin{align}\label{eq:polaron}
    \ket{\Psi(\boldsymbol{{f}})} = \frac{1}{\sqrt{2}} \left(\ket{\uparrow,\boldsymbol{f}} - \ket{\downarrow,-\boldsymbol{f}} \right),
\end{align}
where $\ket{ \boldsymbol{f}} = \exp[ \sum_k (f_k b_k^\dagger - f_k^* b_k)]\ket{0}$ are coherent states of the bath, with the displacements $f_k = -g_k/{2(\omega_q+\omega_k)}$. The residual qubit population in the polaron state is \begin{equation}P_+= \left(1-e^{-2 \sum_k \abs{f_k}^2}\right).\label{eq:respop}\end{equation} It is shown in Fig.~\ref{fig:fig1} by the crosses, which can be seen to be in excellent agreement with the exact numerical results. 

\begin{figure}[t!]
\includegraphics[width=\columnwidth]{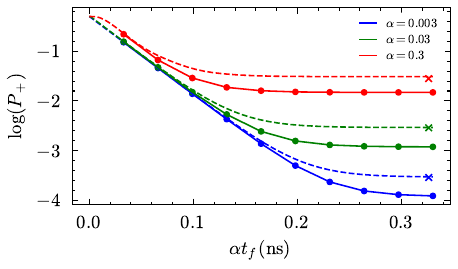}
\caption{Excited state population $P_+$ of the qubit at the end of a reset protocol of duration $t_f$ for coupling strengths $\alpha=0.3$ (top two curves),  $0.03$ (middle two curves),  and $0.003$ (lower two curves). Dashed lines: protocol with a constant qubit splitting. Solid lines and points: protocols obtained by numerical optimization of the time-dependent splittings $\omega_q(t)$. Crosses: predictions of the polaron ansatz.}
\label{fig:fig1}
\end{figure}

\section{Optimal frequency control}
\label{sec:optfreqcont}

To investigate whether the residual excitations can be reduced using a time-dependent qubit Hamiltonian we have implemented numerical optimal control with the process tensor~\cite{butler2024,fux2024}  to minimize the cost function 
\begin{align}
    Z(\rho_f) = 1-\Tr[\rho_f \rho_T].\label{eq:infid}
\end{align}
This is the infidelity between the final density matrix of the system after a protocol of duration $t_f$, $\rho_f = \rho_S(t_f)$, and a target density matrix $\rho_T$. The target is the ground state of the qubit $\rho_T = \ket{-}\bra{-}$. The control parameters, which are varied to minimize Eq.~\eqref{eq:infid}, are the qubit frequencies, $\omega_q(t_n)$ at the discrete times $\{t_n\}$ in the process tensor. We have also considered Hamiltonians containing terms proportional to $\sigma_y$ and $\sigma_z$, but find they remain negligible during optimization. We use the backpropagation method~\cite{butler2024} to obtain the gradient of the cost function with respect to the control parameters, which is used in a numerical minimization of the infidelity with the L-BFGS-B algorithm~\cite{fletcher_practical_2000,virtanen_scipy_2020}. We incorporate bounds on the frequency, constraining it to a reasonable range for a transmon, $\omega_q(t)/2\pi \in  [4,7]$ GHz \cite{krantz2019}.

The points in Fig.~\ref{fig:fig1} show the final excited state populations for the optimized protocols, with each point corresponding to an independent optimization for a given duration $t_f$ and coupling. For short durations $t_f$ the populations are the same as for the constant splitting (dashed lines) because in this regime the qubit does not fully complete the dissipative part of the process, and the constant splitting is already optimal. However, we see that the optimized protocol begins to improve on the constant one once the duration reaches the saturation regime of the latter. The improvement increases with protocol duration until it too saturates; in that regime, the improvement is by a similar factor, $\approx 10^{-0.5}$, for all the cases shown. 
State-of-the-art qubit reset can achieve populations of order $10^{-3}$ in a time of the order of $100$ ns \cite{tuorila2019,zhou2021,magnard2018}, similar to what we obtain for $\alpha = 0.003$.
In the following, however, we analyze the case with $\alpha = 0.03$, which is in the same weak coupling regime but allows for a clearer visualization of the resulting dynamics. 

The time-dependence of the population, $P_+(t)$, and the optimal protocol, $\omega_q(t)$, are shown in Fig.~\ref{fig:fig2}, for $\alpha=0.03$ and $t_f = 11$ ns. In line with the discussion above the optimal protocol, in Fig.~\ref{fig:fig2}(b), is the same as the time-independent one at the beginning of the process, as the qubit dissipates and the population, in Fig.~\ref{fig:fig2}{(a)}, decays exponentially. However, as the relaxation completes and the polaron begins to forms the optimal protocol develops oscillations, whose amplitude increases until it saturates the bounds at the end of the protocol. Their frequency is equal to the constant initial value of the qubit frequency, $\omega_q^0$, as shown in Fig.~\ref{fig:fig2}(c). The excited state population similarly presents increasing oscillations until the protocol is stopped in a minimum of the oscillation in order to maximize the fidelity, which is improved by approximately one-half of an order of magnitude compared with the constant protocol. Interestingly, rather than avoiding the formation of the polaron, the optimal protocol found allows the polaron to form, and then appears to destroy it.

\subsection{Mechanism of polaron destruction}

\begin{figure}[t!]
\includegraphics[width=\columnwidth]{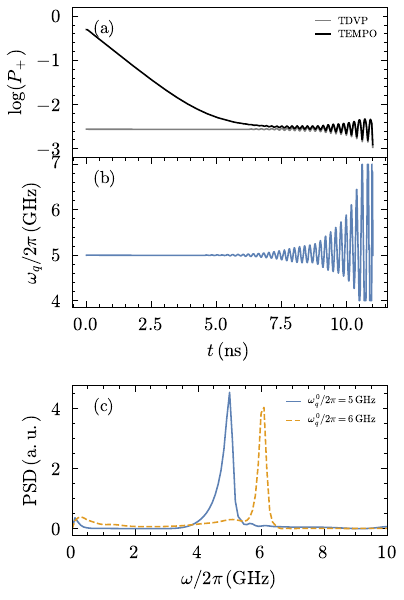}
\centering
\caption{(a) Dynamics of the qubit population for the optimized $\omega_q(t)$ with $\alpha=0.03$ and $t_f=11$ ns. Black: numerical results using OQuPy. Grey: TDVP approximation of the dynamics starting from the polaron state. (b) Time-dependent frequency control, $\omega_q(t)$, found by the optimizer. (c) Power spectrum of the optimal protocol, $\omega_q(t)$, showing that the oscillation frequency matches the initial constant value $\omega_q^0$. }
\label{fig:fig2}
\end{figure}

We now explore how the optimal control destroys the polaron state that forms after the initial dissipative stage. To do this we compute the dynamics of the environment and system-environment correlations. While this can be done using the process tensor~\cite{shortpaper,gribben2022a}, we instead use an approach based on the polaron ansatz. As detailed in~\cite{shortpaper}, we promote the displacements $f_k$ to time-dependent variables and use the time-dependent variational principle (TDVP) \cite{Yao2013, Bera2016, Wang2016,Chen2018,Zhao2022,Hackl2020} to obtain the equations of motion (EOM)
\begin{align}\label{eq:eomold}
    \dot{f}_k = i f_k (\omega_q(t) e^{-2\sum_k \abs{f_k}^2} + \omega_k)  + i\frac{1}{2}g_k.
\end{align} Since we are in the weak coupling regime, where displacements are small, we can neglect the exponential term, removing the coupling between the modes. We write the time-dependent qubit frequency in terms of constant and time-dependent parts, $\omega_q(t)=\omega_q^0+\Delta(t)$, and define renormalized bath frequencies $\omega_k^\prime=\omega_k+\omega_q^0$, to get \begin{align}  
\dot{f}_k = i f_k (\Delta(t)  + \omega_k^\prime)  + i\frac{1}{2}g_k.\label{eq:eom}
  \end{align} These EOM are those of a set of harmonic oscillators, with time-dependent frequencies and an additional driving force from the coupling~\footnote{The additional force becomes time-dependent when a time-dependent qubit-environment coupling is considered, see ~\cite{shortpaper}}. We note that Eqs.~\eqref{eq:eom} are independent of any $\sigma_y$ or $\sigma_z$ terms in the qubit Hamiltonian, consistent with the observation that such terms do not affect the reset process.

\begin{figure*}[t]
\includegraphics[width=2.0\columnwidth]{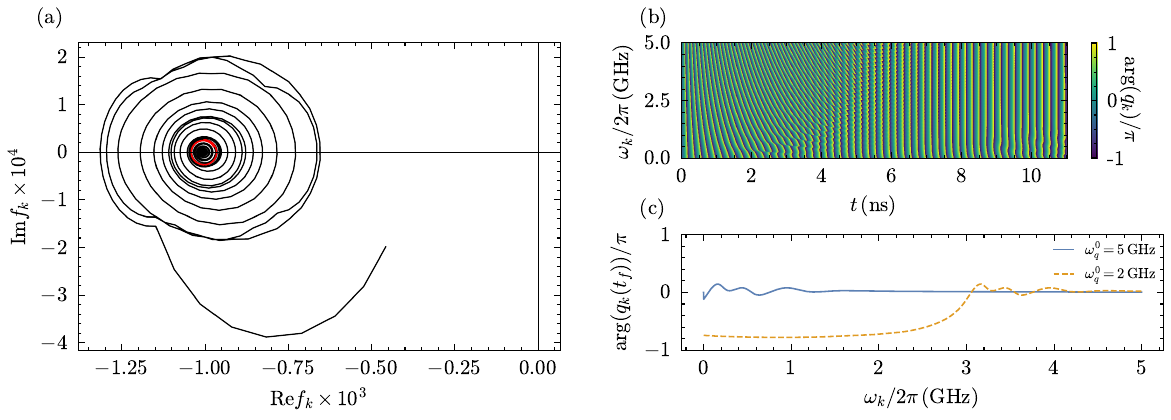}
\centering
\caption{(a) Complex displacement $f_k(t)$ in the polaron state for a bath mode with frequency $\omega_k =  \omega_C/2$, for $\alpha = 0.03$, and the optimal frequency control shown in Fig.~\ref{fig:fig2}(b). Increasing oscillations in the control result in oscillations in the displacement which approach zero. In red we highlight one period of the oscillations, tracing an ellipse in the complex plane with an aspect ratio of $1.45 \approx \omega_k'/\omega_q^0$. (b) Time-dependent phases of the oscillator displacements from equilibrium, $q_k(t) = f_k(t) - f_{k0}$. The optimal frequency control rephases the oscillators such that at final protocol time all oscillators approach the ground state simultaneously. (c) Phase of the oscillators for the optimal protocol at the second to last minima in Fig.~\ref{fig:fig2}(a), while the weak driving assumption is still valid. In a dashed line we have shifted $\omega_q^0$ from the value in the optimal protocol to showcase the predicted $\pi$ phase shift in the oscillations across the resonance. }
\label{fig:oscillator_dynamics_exp}
\end{figure*}

Solving Eqs.~\eqref{eq:eom} for the optimal protocol, starting from the polaron state with $f_k=f_{k0}=-g_k/(2\omega_k^\prime)$, gives the results shown in Figs.~\ref{fig:fig2}(a) and ~\ref{fig:oscillator_dynamics_exp}. The grey curve in Fig.~\ref{fig:fig2}(a) shows the residual population, which agrees with the TEMPO  results (black), demonstrating the validity of the TDVP. The displacements of an oscillator with $\omega_k =  \omega_C/2$ are shown in Fig.~\ref{fig:oscillator_dynamics_exp}(a). The increasing oscillations in the optimal protocol result in the displacement tracing approximate ellipses in the complex plane, with growing amplitude, which leaves the oscillator near to its ground state, $f_k = 0$, at the end of the protocol. In Fig.~\ref{fig:oscillator_dynamics_exp}(b) we show the dynamics of the phases of all the oscillators with respect to their equilibrium points, $\arg(f_k-f_{k0})$. Initially the bath modes oscillate with different frequencies, but when the qubit splitting starts oscillating, the bath modes rapidly rephase. This ensures that the bath modes approach their ground states together at the end of the protocol, minimizing the total bath excitation, $\sum_k |f_k|^2$, and hence the residual population, Eq.~\eqref{eq:respop}. From the point of view of the polaron ansatz, the optimal protocol not only steers the dynamics of the system, but also that of the bath and the system-bath correlations. This is much more difficult than conventional optimal control, where only the dynamics of the system is involved, as it involves an infinite number of degrees of freedom. 

While more complex approaches, such as multipolaron expansions~\cite{bera2014} or Davydov ansatzes~\cite{Yao2013,Chen2018,Zhao2022}, can improve on the accuracy of the polaron form Eq.~\eqref{eq:polaron} in some cases, they are not needed in the weak coupling regime considered here. Furthermore, the polaron ansatz results in the simple Eqs.~\eqref{eq:eom}. These can be solved analytically when the time-dependent frequency shift is small,  $\Delta \ll \omega_k'$ and we have, to first order,
\begin{align}
    f_k(t,\Delta) = f_{k}(t,0) + \Delta \frac{\partial f_k (t,\Delta)}{\partial \Delta}\bigg \lvert_{\Delta=0}.
\end{align}
Taking the unperturbed solution to be $f_k(t,0)=f_{k0}$, and defining the deviation
\begin{align}
    q_k =& f_k(t,\Delta) - f_{k0} \nonumber \\
    \approx & \Delta \frac{\partial f_k (t,\Delta)}{\partial \Delta}\bigg \lvert_{\Delta=0},
\end{align}
gives
\begin{align}
    \dot{q}_k = iq_k\omega_k' + i\Delta(t) f_{k0}.\label{eq:perturbeom}
\end{align} This is the equation of a harmonic oscillator with a fixed frequency and a time-dependent driving.  

To take the environment to its ground state we need all the bath modes to approach $q_k = -f_{k0}$. This involves maximizing displacement from the equilibrium while keeping all oscillators in phase. The former is achieved by a sinusoidal drive but it is not trivial to understand which frequency would be optimal since we have a continuum of bath frequencies. To study this we approximate the time-dependent contribution as $\Delta(t) = h(t)\cos(\Omega t)$, where $h(t)$ is a slowly varying envelope function, and consider the solution to the equation above,
\begin{align}
    q_{k}(t) =& \frac{i   f_{k0}}{2} e^{i\omega_k' t}  \int_0^t dt' \, h(t) e^{-i( \Omega + \omega_k') t'} \nonumber\\
    &+ \frac{i   f_{k0}}{2} e^{i\omega_k' t}  \int_0^t dt' \, h(t) e^{-i(- \Omega + \omega_k') t'}.
\end{align}
Assuming $h(0) = 0$ an asymptotic expansion of the oscillatory integral, to first order, leads to
\begin{align}\label{eq:tdvp}
    q_k(t) =& i\frac{  h(t) f_{k0}}{2}\left[ \frac{1}{\Omega+\omega_k}e^{-i\Omega t}+\frac{1}{-\Omega+\omega_k}e^{i\Omega t}\right] \nonumber \\
    =&i\frac{  h(t) f_{k0}}{(\omega_k'^2-\Omega^2)}\left[\omega_k' \cos(\Omega t) + \Omega i \sin(\Omega t) \right].
\end{align}
This approximation is only valid away from the resonance, where the solution diverges. Nonetheless, the equation above displays the main features of the dynamics. It shows that all bath modes oscillate with the frequency of the drive once it kicks in, and we can identify Eq.~\eqref{eq:tdvp} as the equation of an ellipse in the complex plane. Another important feature, as expected from the harmonic oscillator Eq.~\eqref{eq:perturbeom}, is the $\pi$ phase shift between oscillators with frequencies below and above that of the oscillations in the drive, $\Omega$. Thus any $\Omega>\omega_q^0$ will result in some bath modes being near the maximum of their displacement where others are near the minimum, resulting in a lower fidelity. To demonstrate this we show, in Fig.~\ref{fig:oscillator_dynamics_exp}(c), the final phase of the oscillators, $\arg(q_k(t))$, for the optimal protocol compared with that for the same protocol with a different $\omega_q^0$. While the former shows a relatively homogeneous phase distribution the latter displays a $\pi-$shift as predicted by Eq.~\eqref{eq:tdvp}. Therefore, in order to maximize the $q_k$ while keeping them in phase the optimized solution has
\begin{align}
    \Omega \approx& \min\{\omega_k'\} \nonumber\\
    =& \omega_q^0,
\end{align}
as seen in Fig.~\ref{fig:fig2}(c). 

\section{Tailoring the environment}
\label{sec:tailoring}

\begin{figure}[t]
\includegraphics[width=\columnwidth]{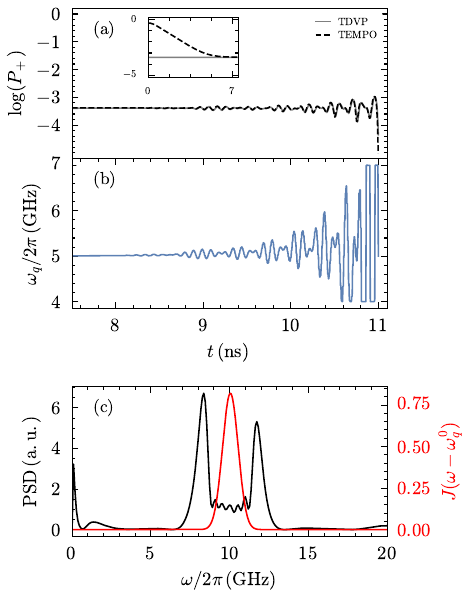}
\centering
\caption{(a) Dynamics of the qubit population for the optimal protocol with a filtered bath at the end of the process. The inset shows the initial dynamics. Black (grey) curve shows the results from TEMPO (TDVP).  (b) Time-dependent frequency control found by the optimizer which oscillates with two frequencies. (c) Power spectrum of the optimal protocol showing two dominant frequencies to either side of the center of the distribution of shifted bath frequencies (red). $\alpha = 0.013, \omega_C/2\pi = 5$ GHz.}
\label{fig:filtered}
\end{figure}

In the reset process above the requirements of reset speed and fidelity compete due to the presence of two different processes. Dissipation occurs with rate  $\propto J(\omega_q(t))$, so that fast reset requires a large system-environment coupling, whereas the polaron formation limits the fidelity according to  $P_+ \approx \sum_k \abs{f_k}^2$, suggesting a small coupling. However, one can improve on this situation by tailoring the spectral density to maximize dissipation while suppressing the entanglement with the bath. One way to do this would be to filter the environment to limit the range of frequencies that interact with the qubit to those near $\omega_q$. Such filters are commonly used for superconducting qubits to avoid unwanted dissipation to readout resonators~\cite{reed2010,yan2023}, and would already improve the fidelity for the standard protocol by reducing the strength of the polaron. Furthermore,  we would expect such filtering to improve the relative performance of the optimal protocol, by reducing the spread of frequencies in the bath and so making it easier to control. 

\begin{figure}[t]
\includegraphics[width=\columnwidth]{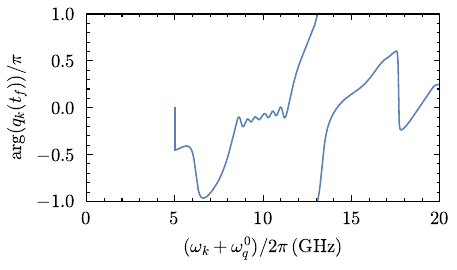}
\centering
\caption{Phase profile of the oscillators at the final time for the optimal protocol with a filtered bath, as a function of the shifted bath frequency $\omega_k' = \omega_q^0 + \omega_k$. In the frequency region of the filter, centered on $\omega_q^0+\omega_C = 10$ GHz, the phase is approximately constant, but varies strongly outside this region.}
\label{fig:phase_filtered}
\end{figure}

To explore this we modify the spectral density to include a Gaussian filter, with center frequency $\omega_C$ and standard deviation $\sigma$, 
\begin{align}
    J(\omega) =&2\alpha \omega e^{-(\omega-\omega_c)^2/2\sigma^2}.
\end{align} We choose $\alpha = 0.013$, such that the decay rate is similar to that without the filter studied above. We show the resulting optimal protocol, and the time-dependent residual population, in Fig.~\ref{fig:filtered}. There is an improvement of the fidelity for the standard protocol, as expected, but more noticeable is that optimization now brings a larger improvement, of more than one order-of-magnitude, to give a residual population of $10^{-5}$. The optimal control is similar to that for the unfiltered bath, with a constant initial value of the frequency until the polaron forms, and growing oscillations which act to destroy it. The main difference with respect to the unfiltered case is the frequencies of oscillation in the control. As can be seen in the power spectrum of the control $\omega_q(t)$, in Fig.~\ref{fig:filtered}(c), there are now two frequencies in the oscillation, one on each side of the center peak of the distribution of renormalized bath frequencies, $\omega_q^\prime=\omega_q^0+\omega_C$. 

The dynamics of the bath modes (not shown) is similar to the Ohmic case shown in Fig.~\ref{fig:oscillator_dynamics_exp}, with rephasing of the oscillators and displacements that grow in the complex plane approaching the ground state. Therefore, while a single driving frequency at $\omega_q^0+\omega_C$ might be favorable to drive the amplitude of the bath modes closer to $f_k=0$, this is not optimal because of the same $\pi-$shift discussed above. Instead the optimal protocol displays two frequencies just outside of the Gaussian, which we argue serve to homogenize the phase of the oscillations around the equilibrium point. In Fig.~\ref{fig:phase_filtered} we show the final phase of the bath modes which indeed show an approximately uniform phase in the region of the filter.

\section{Multi-level transmon}
\label{sec:multilevel}

So far we have treated the qubit as a two-level system. This is the ideal case, but in practice transmons are anharmonic oscillators. Although their anharmonicity allows them to be approximated by their two lowest eigenstates, leakage to higher levels is possible, and it is in fact a notable source of error~\cite{babu2021, mcewen2021}. We therefore must analyze how polaron formation affects higher levels and whether the optimal control found is still valid outside of the ideal case. Accounting for the higher levels, the full Hamiltonian is given by
\begin{align}
    H =& \omega_q a^\dagger a +\frac{\alpha_A}{2} a^{\dagger \, 2}a^2+ \sum_k \omega_k b^\dagger_k b_k \nonumber \\
    &+\frac{1}{2}(a+a^\dagger)\sum_k g_k (b_k+b^\dagger_k) ,
\end{align}
where $\alpha_A$ is the anharmonicity and there is a dipole-dipole interaction with the bath. Note that truncating to a two-level system, $a^\dagger a = \sigma_x+1/2$ and $(a+a^\dagger)=\sigma_z$, reduces this Hamiltonian to the spin-boson Hamiltonian in Eq.\eqref{eq:sb}. 

\begin{figure}[t]
\includegraphics[width=\columnwidth]{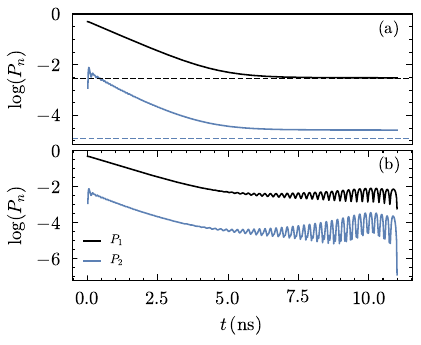}
\centering
\caption{Populations of the first two excited states of a transmon, modelled as a truncated anharmonic oscillator, during reset with a (a) constant harmonic oscillator frequency and (b) the optimized time-dependent frequency shown in Fig.~\ref{fig:fig2}(b). The dashed lines in (a) show the predictions of the multilevel polaron ansatz. The oscillator is truncated to its lowest 5 levels. The population of the higher excited states shows similar behavior.}
\label{fig:multilevel}
\end{figure}

To approximate the polaron ground state, we extend the variational approach employed for the spin-boson model (see Appendix \ref{appendix:multilevel} for details). We use an ansatz where the bath modes are displaced according to the system-bath coupling,
\begin{align}
    \ket{\Psi(\vec{f})} = e^{(a+a^\dagger)\sum_k ( f_k b_k^\dagger - f_k^* b_k)}\ket{0,0},
\end{align}
and find that the energy is minimized for the displacements
\begin{align}
    f_k = \frac{-g_k}{2(\omega_q + 3\alpha_A \sum_q \abs{f_q}^2 + \omega_k)},
\end{align}
similarly to the two-level case, with both $\alpha_A$ and $\sum_q \abs{f_q}^2$ being small parameters. The populations of the transmon in the polaron ground state are
\begin{align}
    P_n   =&\bra{\vec{0}}   \frac{(i\hat{B})^{2n}}{n!} e^{\hat{B}^2}  \ket{\vec{0}}\nonumber \\
    =&  \sum_{p=0}^\infty \frac{(-1)^p}{n!p!}\left(\sum_k \abs{f_k}^2 \right)^{n+p} (2(n+p)-1)!!   ,
\end{align}
where $\hat{B} = \sum_k f_k(b_k^\dagger-b_k) $. In Fig.~\ref{fig:multilevel}(a) we compare TEMPO results for the dynamics with a constant frequency, treating the transmon as a $5$-level system, with those from the polaron ansatz. We find that the simple polaron ansatz is less accurate in describing the higher excited states in the qubit, but remains a good approximation. In Fig.~\ref{fig:multilevel}(b) we compute the dynamics under the optimal control protocol shown in Fig.~\ref{fig:fig2}(b). Note that in a flux-tunable transmon, only $\omega_q$ is controlled, while the anharmonicity $\alpha_A$ remains constant~\cite{krantz2019}. We find that the effect on the first excited state is the same as in the two-level case. The optimal control also reduces significantly the population of the higher excited states of the qubit, which remain orders of magnitude below the first excited state. 

\section{Conclusions}
\label{sec:concl}

Qubit reset shows how the high accuracies sought in quantum technology can demand treatments beyond the Born-Markov approximation even for nominally weak couplings. Analyzing reset of a transmon qubit using a spin-boson model, treating it using the exact PT-TEMPO algorithm, shows that limitations arise due to system-bath correlations which correspond to the formation of the polaron. Implementing optimal control within this algorithm allows us to show that these limitations can be overcome by time-dependent controls of the qubit. The dynamics under these controls, investigated using the time-dependent variational principle, shows that they destroy the polaron by driving phase-coherent oscillations of the bath modes about equilibrium in the correlated system-environment state. In conjunction with filtering of the bath, to improve its controllability and weaken the polaron, this gives excited state populations $P_{+}\sim 10^{-5}$ at zero temperature, in a time $\sim 10$ ns. This is comparable to the thermal populations at realistic transmon temperatures and reaches the levels required for error correction, while being an order-of-magnitude faster than the state-of-the-art. We find similar or better performance for a time-dependent coupling~\cite{shortpaper}, with simpler protocols which may therefore be more robust. As well as its relevance to qubit reset our method and results could be useful for improving the performance of other devices, such as single-photon sources, which are limited by polaron formation. It would also be interesting to explore the impact of these polaron effects, and the possibility of their control, for the thermodynamics of heat transfers~\cite{popovic2021}, work statistics~\cite{shubrook_numerically_2025}, and the Landauer bound~\cite{maruyama2009}, and the implications for quantum thermal machines. More generally, our work extends control theory to demonstrate manipulation of the full many-body state of an open quantum system, including the environment and system-environment correlations, enabling new approaches to studying and exploiting quantum dynamics.

\begin{acknowledgments}
We acknowledge funding from Taighde \'Eireann -- Research Ireland (21-FF-P/10142). 
\end{acknowledgments}

In order to meet institutional and research funder open access requirements, any accepted manuscript arising shall be open access under a Creative Commons Attribution (CC BY) reuse license with zero embargo.

Code and data supporting this article are openly available~\cite{datalong}. 

\appendix

\section{Analytical calculation of the influence functional blocks.}\label{appendix:influence}

The TEMPO algorithm relies on decomposing the influence functional into a tensor network of blocks computed from the object
\begin{align}
    \eta_k =& \int_{k \delta t}^{k\delta t+\delta t}dt'\int_{0}^{\delta t} dt''\, C(t'-t'')\nonumber \\
    =& \eta(k\delta + \delta t) - 2\eta(k\delta t)+\eta(k\delta - \delta t),
\end{align}
where 
\begin{align}
    C(t) = \int_0^\infty d\omega \, J(\omega) \left[ \coth(\frac{\omega}{2T})\cos(\omega t) - i \sin(\omega t) \right],
\end{align}
is the correlation function and $\eta_k$ is expressed in terms of 
\begin{align}
    \eta(t) =& \int_0^\infty d\omega \, \frac{J(\omega)}{\omega^2}\left[ \coth(\frac{\omega}{2T})[\cos(\omega t)-1] \right.\nonumber\\
    & \left. \vphantom{\coth(\frac{\omega}{2T})} - i[\sin(\omega t)-\omega t]\right] .
\end{align}
For the parameters relevant in qubit reset one must simulate a number of timesteps of order $10^3-10^4$ to be able to resolve the correlations in the bath. Evaluating these numerically is time consuming, in part due to their oscillatory nature. To circumvent this problem and to improve the efficiency of the calculation we evaluate these integrals analytically in terms of special functions. For the case relevant here, an Ohmic bath with an exponential cutoff, $J(\omega) = 2\alpha \omega \exp(-\omega/\omega_C)$, at zero temperature we have
\begin{align}
    \eta(t) =& 2\alpha \int_0^\infty d\omega \frac{1}{\omega}e^{-\omega/\omega_C} \left(e^{-i\omega t} -1 \right) \nonumber\\
    &+ 2\alpha t i \int_0^\infty d\omega e^{-\omega/\omega_C} \nonumber\\
    =& 2\alpha\left(-\log(i\omega_C t +1) + i t \omega_C \right).
\end{align}
At finite temperature we must separate the integral into two converging terms, consider first
\begin{align}
    I_1 =& \int_0^\infty d\omega \, \frac{J(\omega)}{\omega^2}\coth(\frac{\omega}{2T})[\cos(\omega t)-1] \\
    =& 2\alpha \Re \int_0^\infty d\omega \, \frac{e^{-\omega/\omega_C}}{\omega}\frac{1+e^{-\omega/T}}{1-e^{-\omega/T}}\left( e^{i\omega t}  -1 \right) \nonumber \\
    =& 2\alpha \Re \int_0^\infty dx \, \frac{e^{-x(T/\omega_C+1/2)}}{x}\frac{e^{x/2}+e^{-x/2}}{1-e^{-x}}\left( e^{ix T t}  -1 \right) \nonumber \\
    =& 2\alpha \sum_{k_j=\pm} k_2 \Re \int_0^\infty dx \, \frac{e^{-x(T/\omega_C+1/2+k_1/2)}}{x(1-e^{-x})}\left( e^{ix T t}  -1 \right). \nonumber
\end{align} 
This integral can be expressed in terms of 
\begin{align}
     &\int_0^\infty dx \, x^{-1}\frac{e^{-(a+ib)x}-e^{-ax}}{1-e^{-x}} \\
     =&-\int_0^\infty dx \int_a^{a+ib} dz\, \frac{e^{-zx}}{1-e^{-x}} \nonumber \\
     =&- \int_a^{a+ib} dz\, \left(\int_0^\infty dx\, x^{-1}e^{-x} - \psi(z) \right) \nonumber \\
     =&-ib \int_0^\infty dx\, x^{-1}e^{-x} + \int_a^{a+ib} dz\, \psi(z) \nonumber \\
     =&-ib \int_0^\infty dx\, x^{-1}e^{-x} + \ln \Gamma(a+ib) -\ln \Gamma(a) \nonumber ,
\end{align}
where we used the definition of the digamma function, $\psi(z) = d \ln \Gamma(z)/dz$, as well as its integral representation. The first term is divergent, but it will vanish in $I_1$ when the real part is taken. Using the result above we have
\begin{align}
    I_1  =&  2\alpha \sum_{k_j = \pm} k_2 \Re  \ln \Gamma \left(\frac{T}{\omega_C} + \frac{k_1+1}{2} + iTt\frac{k_2+1}{2}\right),
\end{align}
while the second term results in
\begin{align}
        I_2 =& -i\int_0^\infty d\omega \, \frac{J(\omega)}{\omega^2} [\sin(\omega t)-\omega t] \nonumber\\
        =& -2i\alpha(\arctan(\omega_C t) - \omega_C t).
\end{align}

\section{Polaron formation in multi-level transmon}\label{appendix:multilevel}

The transmon is modelled as an  anharmonic oscillator,
\begin{align}
    H_Q = \omega_q a^\dagger a + \frac{\alpha_A}{2} a^{\dagger \, 2} a^2.
\end{align}
We consider a weak dipole-dipole interaction with a bath of harmonic oscillators
\begin{align}
    H_{QB} =&\sum_k \omega_k b^\dagger_k b_k + \frac{a+a^\dagger}{2}\sum_k g_k (b_k+b^\dagger_k) .
\end{align}
Similarly to the two-level case we consider an ansatz where the bath modes are displaced according to the system coupling operator,
\begin{align}
    \ket{\Psi(\vec{f})} = e^{(a+a^\dagger)\sum_k ( f_k b_k^\dagger - f_k^* b_k)}\ket{0,0}.
\end{align}
This operator can be expressed in terms of the displacement operators of the qubit or the bath,
\begin{align}
    \ket{\Psi(\vec{f})} =&D_a(\hat{B})\ket{0,0} \nonumber \\
    =&\prod_k D_{b_k}(f_k \hat{x})\ket{0,0},
\end{align}
where $\hat{B} =\sum_k (f_k b_k^\dagger - f_k^* b_k)$ and $\hat{x} = a+a^\dagger$. This allows us to easily compute the expectation value of the bare Hamiltonian,
\begin{align}
    \bra{\Psi(\vec{f})}a^\dagger a \ket{\Psi(\vec{f})} =& \bra{0,0}(a^\dagger - \hat{B})(a+\hat{B})\ket{0,0} \nonumber \\
    =& -\sum_k \abs{f_k}^2 \bra{0,0}(-b_k b_k^\dagger)\ket{0,0} \nonumber \\
    =& \sum_k \abs{f_k}^2 ,
\end{align}
as well as
\begin{align}
    \bra{\Psi(\vec{f})}&a^{\dagger \,2} a^2 \ket{\Psi(\vec{f})} \nonumber \\
    =& \bra{0,0}(a^\dagger - \hat{B})^2(a+\hat{B})^2\ket{0,0} \nonumber \\
    =& 3\left( \sum_k \abs{f_k}^2\right)^2,
\end{align}
and
\begin{align}
    \bra{\Psi(\vec{f})}b_k^\dagger b_k \ket{\Psi(\vec{f})} =& \bra{0,0}(b_k^\dagger + \hat{x}f_k^*)(b_k+ \hat{x}f_k)\ket{0,0} \nonumber \\
    =&  \abs{f_k}^2 \bra{0,0}\hat{x}^2\ket{0,0} \nonumber \\
    =&  \abs{f_k}^2 .
\end{align}
For the interaction term we have,
\begin{align}
    \bra{\Psi(\vec{f})}&(a+a^\dagger) (b_k + b_k^\dagger) \ket{\Psi(\vec{f})} \nonumber\\
    =& \bra{0,0}(a+a^\dagger)(b_k + \hat{x}f_k + b_k^\dagger + \hat{x}f_k^*)\ket{0,0} \nonumber\\
    =& (f_k + f_k^*) \bra{0,0} \hat{x}^2\ket{0,0} \nonumber\\
    =& (f_k + f_k^*).
\end{align}
Altogether the expectation value of the Hamiltonian becomes
\begin{align}
    \bra{\Psi(\vec{f})} H \ket{\Psi(\vec{f})} =& \omega_q \sum_k \abs{f_k}^2 +\frac{3\alpha_A}{2} \left( \sum_k \abs{f_k}^2\right)^2 \nonumber \\
    &+ \sum_k \omega_k \abs{f_k}^2 + \frac{1}{2}\sum_k g_k (f_k + f_k^*),
\end{align}
which is minimized for 
\begin{align}
    f_k = \frac{-g_k}{2(\omega_q + 3\alpha_A \sum_q \abs{f_q}^2 + \omega_k)}.
\end{align}

We now compute the populations of the transmon by first considering the amplitude
\begin{align}
    \bra{n}\ket{\Psi(\vec{f}} =&  \bra{n} e^{a^\dagger \hat{B}}e^{a \hat{B}}e^{\hat{B}^2/2} \ket{0} \otimes \ket{\vec{0}} \nonumber \\
    =&e^{\hat{B}^2/2}  \bra{n} \sum_m \frac{(a^\dagger \hat{B})^m}{m!}\ket{0} \otimes \ket{\vec{0}} \nonumber \\
    =&e^{\hat{B}^2/2}  \frac{(\hat{B})^n}{\sqrt{n!}} \otimes \ket{\vec{0}},
\end{align}
resulting in the population
\begin{align}
    P_n =& \bra{\vec{0}}   \frac{(- \hat{B})^n}{\sqrt{n!}}       e^{\hat{B}^2}  \frac{( \hat{B})^n}{\sqrt{n!}}  \ket{\vec{0}} \nonumber \\
    =&\bra{\vec{0}}   \frac{(i\hat{B})^{2n}}{n!} e^{\hat{B}^2}  \ket{\vec{0}} .
\end{align}

To compute this expectation value, first consider,
\begin{align}
    r_n =& \bra{\vec{0}}  \hat{B}^{2n} \ket{\vec{0}} \nonumber \\
    =&\sum_k f_k \bra{\vec{0}}  \hat{B}^{2n-1} b_k^\dagger \ket{\vec{0}} \nonumber \\
    =&-(2n-1)\sum_k \abs{f_k} \bra{\vec{0}}  \hat{B}^{2n-2} \ket{\vec{0}} \nonumber\\
    =&-(2n-1)\sum_k \abs{f_k}^2 r_{n-1}.
\end{align}
where we have used the commutator $[B^N,b_q^\dagger] = -N f_q^*  B^{N-1}$. Together with $r_0 = 1$, solving recursively we obtain
\begin{align}
    r_n =& \left(-\sum_k \abs{f_k}^2 \right)^n \prod_{m=1}^n (2m-1) \nonumber\\
    =&\left(-\sum_k \abs{f_k}^2 \right)^n (2n-1)!! ,
\end{align}
which can be used to calculate the populations
\begin{align}
    P_n =& \frac{(-1)^n}{n!}\sum_{p=0}^\infty \frac{r_{n+p}}{p!} .
\end{align}

\bibliography{landreset.bib,addmaterials.bib}

\end{document}